\documentclass[]{article}
\usepackage{lmodern}
\usepackage{amssymb,amsmath}
\usepackage{ifxetex,ifluatex}
\usepackage{fixltx2e} 
\ifnum 0\ifxetex 1\fi\ifluatex 1\fi=0 
  \usepackage[T1]{fontenc}
  \usepackage[utf8]{inputenc}
\else 
  \ifxetex
    \usepackage{mathspec}
  \else
    \usepackage{fontspec}
  \fi
  \defaultfontfeatures{Ligatures=TeX,Scale=MatchLowercase}
\fi
\IfFileExists{upquote.sty}{\usepackage{upquote}}{}
\IfFileExists{microtype.sty}{%
\usepackage{microtype}
\UseMicrotypeSet[protrusion]{basicmath} 
}{}
\usepackage[margin=1in]{geometry}
\usepackage{hyperref}
\hypersetup{unicode=true,
            pdftitle={Shapley Decomposition of R-Squared in Machine Learning Models},
            pdfauthor={Nickalus Redell},
            pdfborder={0 0 0},
            breaklinks=true}
\usepackage{graphicx,grffile}
\makeatletter
\def\maxwidth{\ifdim\Gin@nat@width>\linewidth\linewidth\else\Gin@nat@width\fi}
\def\maxheight{\ifdim\Gin@nat@height>\textheight\textheight\else\Gin@nat@height\fi}
\makeatother
\setkeys{Gin}{width=\maxwidth,height=\maxheight,keepaspectratio}
\IfFileExists{parskip.sty}{%
\usepackage{parskip}
}{
\setlength{\parindent}{0pt}
\setlength{\parskip}{6pt plus 2pt minus 1pt}
}
\providecommand{\tightlist}{%
  \setlength{\itemsep}{0pt}\setlength{\parskip}{0pt}}
\ifx\paragraph\undefined\else
\let\oldparagraph\paragraph
\renewcommand{\paragraph}[1]{\oldparagraph{#1}\mbox{}}
\fi
\ifx\subparagraph\undefined\else
\let\oldsubparagraph\subparagraph
\renewcommand{\subparagraph}[1]{\oldsubparagraph{#1}\mbox{}}
\fi

\let\rmarkdownfootnote\footnote%
\def\footnote{\protect\rmarkdownfootnote}

\usepackage{titling}

  \title{Shapley Decomposition of R-Squared in Machine Learning Models}
    \pretitle{\vspace{\droptitle}\centering\huge}
  \posttitle{\par}
    \author{Nickalus Redell}
    \preauthor{\centering\large\emph}
  \postauthor{\par}
    \date{}
    \predate{}\postdate{}
  
\usepackage{float}
\usepackage{caption}

\title{Shapley Decomposition of R-Squared in Machine Learning Models}

\author{
  {\normalfont
  Nickalus Redell \\
  \texttt{nickalusredell@gmail.com} \\
  August, 2019
  }
}

\begin{document}
\maketitle
\begin{abstract}
In this paper we introduce a metric aimed at helping machine learning
practitioners quickly summarize and communicate the overall importance
of each feature in any black-box machine learning prediction model. Our
proposed metric, based on a Shapley-value variance decomposition of the
familiar \(R^2\) from classical statistics, is a model-agnostic approach
for assessing feature importance that fairly allocates the proportion of
model-explained variability in the data to each model feature. This
metric has several desirable properties including boundedness at 0 and 1
and a feature-level variance decomposition summing to the overall model
\(R^2\). In contrast to related methods for computing feature-level
\(R^2\) variance decompositions with linear models, our method makes use
of pre-computed Shapley values which effectively shifts the
computational burden from iteratively fitting many models to the Shapley
values themselves. And with recent advancements in Shapley value
calculations for gradient boosted decision trees and neural networks,
computing our proposed metric after model training can come with minimal
computational overhead. Our implementation is available in the
\(\texttt{R}\) package \(\texttt{shapFlex}\).
\end{abstract}

\hypertarget{introduction}{%
\subsection{Introduction}\label{introduction}}

The recent emphasis on providing human-interpretable explanations of the
outputs of black-box machine learning models shows little sign of
slowing. The motivations to demystify the often-complex workings of
modern predictive models are many and varied {[}1{]} as is the evidence
of their growing importance: 10s of thousands of recently published
academic articles {[}2{]}, ``right to an explanation'' legislation from
the European Union {[}2{]}, and the popularity of open-source software
packages such as \(\texttt{shap}\) {[}3{]}, \(\texttt{DALEX}\) {[}4{]},
and \(\texttt{iml}\) {[}5{]}. In an effort to summarize this literature
and clarify the discussion, several authors have developed useful
taxonomies and recommendations for machine learning practitioners
interested in model interpretability {[}1{]}, {[}2{]}, {[}6{]}.

One aspect of interpretable machine learning relates to understanding
how changes in predictive model inputs or features relate to changes in
model outputs. Focusing on the methods in this space, the framework
below is useful for orienting our proposal of a new model interpretation
method, yet another version of \(R^2\), amongst the many existing
methods:

\begin{enumerate}
\def\labelenumi{\arabic{enumi}.}
\tightlist
\item
  Algorithm-agnostic to algorithm-specific
\item
  Local to regional {[}7{]} to global
\item
  Feature effects to feature importance {[}6{]}.
\end{enumerate}

Using this outline as a guide, our proposed \(R^2\) metric would be
classified as an (a) algorithm-agnostic, (b) global measure of (c)
feature importance. Taking these properties in turn, our proposed metric
is algorithm-agnostic in the sense that it can be applied to all classes
of linear and non-linear machine learning prediction models. It is
global because it returns one summary statistic for each feature.
Lastly, it is a feature importance metric due to its emphasis on ranking
feature influence as opposed to identifying how a change in a feature's
value affects model predictions (i.e., feature effects).

The wisdom of relying on such blunt measures of feature importance for
useful insights into model behavior has been called into question
{[}8{]}. Certainly, methods based on path analysis and causal modeling,
feature effect visualizations similar to partial dependence plots, and
identifying a minimal subset of model features that, when altered,
dramatically affect a model's predictions {[}9{]} all appear to provide
richer information. These methods notwithstanding, it is not uncommon
for researchers to be called upon to identify the most influential
features in the models that they have built. And because the uses for
such summary information range from (a) ease of communication to (b)
identifying possibilities for interventions to (c) assessing the impact
of concept drift on future model predictions, they have a place in the
practitioner's toolbox.

\hypertarget{r2}{%
\subsection{\texorpdfstring{\(R^2\)}{R\^{}2}}\label{r2}}

\(R^2\) has a long and storied history in statistics as a method for
quantifying the variance in an outcome explained by linear models like
linear and logistic regression {[}10{]}. \(R^2\), also known as the
coefficient of determination, and its extensions have been characterized
in a variety of more or less equivalent ways--e.g., as correlations,
explained variance, explained variation--with substantive differences
depending on whether or not the model has an intercept term {[}11{]}.
When described as a proportion of variance explained metric it takes on
the general formula of

\[R^{2}_{classical} = \dfrac{var_{\hat{y}}}{var_{y}} \tag{1}\]

which is the ratio of the variance in the model-predicted values,
\(\hat{y}\), to the variance in the outcome, \emph{y}. In this context,
\(R^2\) is a measure of overall model fit along a 0 to 1
scale--\(-\infty\) to 1 in degenerate cases--with higher values
indicating better fit.

In contrast to its typical use as a measure of overall model fit,
\(R^2\) has also seen use as a feature importance metric {[}8{]}. A
variety of \(R^2\) feature importance methods have been proposed under
the label of variance decomposition. A select few have the desirable
properties of non-negativity and that the variance in the outcome
explained by each feature adds up to the overall model \(R^2\) {[}8{]}.
And all fall under the category of dispersion-related measures of
feature importance. In contrast to feature importance metrics based on
changes in predictive accuracy {[}12{]}, the goal of dispersion-based
metrics is to assess the extent to which a phenomenon can be
explained--on a 0 to 1 or 0\% to 100\% scale--by knowing the values of
the modeled features.

\hypertarget{r2-for-feature-importance-in-machine-learning}{%
\subsection{\texorpdfstring{\(R^2\) for Feature Importance in Machine
Learning}{R\^{}2 for Feature Importance in Machine Learning}}\label{r2-for-feature-importance-in-machine-learning}}

The aim of our proposal is to take a familiar statistical concept,
\(R^2\), and apply it in the machine learning setting as a robust and
intuitive measure of feature importance. The general approach is to
first calculate an appropriate \(R^2\) metric for the model and then
decompose this variance into feature-level shares or attributions. This
repurposing comes at a cost, though. Namely, with machine learning
models like boosted decision trees and neural networks, the same set of
features can explain an arbitrarily large proportion of the total
variance in an outcome through hyperparameter tuning and overfitting.
This suggests that a purely data-based metric like the classical \(R^2\)
from linear modeling may be less fit for purpose than a model-dependent
approach using, for instance, cross-validation to assess the
\emph{expected} variance of a model's residuals and, by extension, the
feature-level \(R^2\) values.

To produce such a metric, we take inspiration from (a) Gelman et. al's
proposed \(R^2\) metric for Bayesian regression {[}13{]} and (b) a
Shapley-value-based variance decomposition of a model's overall \(R^2\)
value. A brief background of each is provided in the following section.

\hypertarget{related-work}{%
\subsection{Related Work}\label{related-work}}

\hypertarget{gelmans-bayesian-r2}{%
\subsubsection{\texorpdfstring{Gelman's Bayesian
\(R^{2}\)}{Gelman's Bayesian R\^{}\{2\}}}\label{gelmans-bayesian-r2}}

Building off of the survival analysis literature, Gelman et.
al.~{[}13{]} proposed a version of \(R^2\) for overall model fit in
Bayesian regression using a variance decomposition that bounds \(R^2\)
at 0 and 1 by construction. While their Bayesian-motivated reasoning is
different than our's, their general proposal of

\[R^{2}_{adjusted} = \dfrac{Explained \ variance}{Explained \ variance + Residual \ variance}\tag{2}\]

has several useful properties. In addition to the 0 to 1 scaling, this
metric is algorithm-agnostic: It works as well for deep neural networks
as it does for linear regression. It is important, here, to note the
distinction between algorithm-agnostic and model-agnostic. While this
formulation works for any class of prediction model, the resulting
\(R^2\) is based on the variance in the model predictions and not on the
observed variance of the outcome as in the traditional version. Put
another way, the denominator in (2) is not fixed--i.e., it is
model-specific--which leaves us without a single source,
dataset-specific amount of variance that we can claim to have explained
{[}13{]}. The authors address this problem as it relates to Bayesian
inference. This issue of grounding \(R^2_{adjusted}\) as a measure of
overall model fit when comparing similar or nested models, however, is
of less concern when the goal is to decompose \(R^2\) to assess global
feature importance.

\hypertarget{shapley-variance-decomposition-of-r2}{%
\subsubsection{\texorpdfstring{Shapley Variance Decomposition of
\(R^{2}\)}{Shapley Variance Decomposition of R\^{}\{2\}}}\label{shapley-variance-decomposition-of-r2}}

Decomposing a model's \(R^2\) into feature-level shares of explained
variance is not a new concept. One of the more popular approaches
involves sequential testing where a feature or groups of features are
added to a baseline model, the model is retrained, and any incremental
increases in \(R^2\) are attributed solely to the new feature(s)
{[}8{]}. The main drawback, of which there are several, is that if
features are correlated, the order in which they are entered into a
model affects their assigned share of variance explained in this first
come first serve system. Enter, the Shapley value.

More precisely, enter the LMG. LMG is a measure of feature importance
proposed by Lindeman, Merenda, and Gold {[}14{]} that decomposes \(R^2\)
into non-negative, non-order-dependent, feature-level shares of variance
explained whose sum totals the overall \(R^2\). The popular LMG approach
of calculating marginal feature importance by examining changes in
linear model fit across all possible feature groupings entering the
model in all possible orders is, however, equivalent to a Shapley value
calculation {[}15{]}.

Shapley values are a concept from cooperative game theory {[}16{]} that
has recently gained popularity in machine learning as a tool for
interpreting predictions from black-box models. Their utility for
providing hyper-local insight into how each model feature influences a
prediction for a given instance places them in a small but important
class of feature interpretability methods. The work of Lundburg and Lee
{[}3{]} in unifying a variety of seemingly different feature explanation
methods using a Shapley-based framework has highlighted their many
beneficial properties. Indeed, Shapley-based explanations of machine
learning predictions have been referred to as potentially ``the only
method to deliver a full explanation. In situations where the law
requires explainability -- like EU's `right to explanations' -- the
Shapley value might be the only legally compliant method, because it is
based on a solid theory and distributes the effects fairly'' {[}5{]}.

The most relevant property for our proposed metric is that Shapley
values are an additive feature attribution method. As identified in
{[}3{]}, the additive property of Shapley values,

\[\hat{y}_{i} = \phi_{0}+\sum_{f=1}^{F}\phi_{i}^{(f)}\tag{3},\]

indicates that for a given instance, \emph{i}, the sum of the
feature-level attributions, \(\phi_{i}^{(f)}\), and the average
prediction across instances in a dataset, \(\phi_{0}\), equals the model
prediction. This amazing decomposition of a single prediction into its
constituent parts across model features is one of the main goals of
Shapley value analysis in machine learning and is directly related to
existing methods for variance decomposition of \(R^2\) {[}8{]}.

\hypertarget{proposed-r2-feature-importance-metric}{%
\subsection{\texorpdfstring{Proposed \(R^2\) Feature Importance
Metric}{Proposed R\^{}2 Feature Importance Metric}}\label{proposed-r2-feature-importance-metric}}

Our proposed metric takes advantage of the robustness of Gelman et. al's
\(R_{adjusted}^2\) and the additive property of Shapley values to
achieve an \(R^2\) variance decomposition that fairly reflects each
feature's contribution to the model's explanatory power. The logic
behind our proposal is that, for a given feature with a non-zero effect,
removing the marginal contribution of that feature from the model's
predictions--as measured by the feature-level vector of Shapley
values--should decrease model accuracy and increase the variance of the
residuals. Features can then be ranked based on the extent to which
their removal increases residual variance, with larger increases in
residual variance indicating more important features.

To introduce notation and summarize the calculation, the proposed metric
is calculated as follows: For each model feature, \emph{f} in 1 to
\emph{F}, a new prediction can be made for outcome \(y\) for each
instance, \emph{i} in 1 to \emph{N}, by subtracting the feature's
instance-level Shapley value, \(\phi^{(f)}_{i}\), from \(\hat{y}_{i}\)
to get \(\hat{y}^{(f)}_{i_{shap}}\) followed by a simple variance
decomposition and normalization to ensure that the feature-level
\(R^{2(f)}\) values sum to the overall model \(R^{2}\). The following
steps outline the \(R^{2}\) calculation in detail.

\begin{quote}
\begin{enumerate}
\def\labelenumi{\arabic{enumi}.}
\tightlist
\item
  Compute the baseline \(R^{2}\) value across all instances of interest
  with all model features with
\end{enumerate}
\end{quote}

\begin{quote}
\[R^{2}_{baseline} = \dfrac{var_{\hat{y}}}{var_{\hat{y}} + var_{res}} \tag{4}\]
\end{quote}

\begin{quote}
\begin{quote}
where \(var_{\hat{y}}\) is the variance of the model's predictions and
\(var_{res}\) is the variance of the model residuals (\(y - \hat{y}\)).
\end{quote}
\end{quote}

\begin{quote}
\begin{enumerate}
\def\labelenumi{\arabic{enumi}.}
\setcounter{enumi}{1}
\tightlist
\item
  For each model feature \emph{f} in \emph{F}, compute the
  Shapley-modified prediction \(\hat{y}_{shap}\) with
\end{enumerate}
\end{quote}

\begin{quote}
\[\hat{y}^{(f)}_{i_{shap}} = \hat{y}_{i} - \phi^{(f)}_{i} \tag{5}.\]
\end{quote}

\begin{quote}
\begin{quote}
As a vectorized operation, the input matrix would be \(N \times F\) rows
with each instance, \emph{i}, receiving one Shapley-altered prediction
per feature.
\end{quote}
\end{quote}

\begin{quote}
\begin{enumerate}
\def\labelenumi{\arabic{enumi}.}
\setcounter{enumi}{2}
\tightlist
\item
  For each model feature \emph{f} in \emph{F}, compute the feature-level
  \(R^{2}\) with
\end{enumerate}
\end{quote}

\begin{quote}
\[R_{shap}^{2(f)} = \dfrac{R_{baseline}^{2} - min(\dfrac{var_{res_{baseline}}}{var_{res_{shap}}^{(f)}}, 1) \times R_{baseline}^{2}}{\sum_{f=1}^{F}R_{baseline}^{2} - min(\dfrac{var_{res_{baseline}}}{var_{res_{shap}}^{(f)}}, 1) \times R_{baseline}^{2}} \times R_{baseline}^{2} \tag{6}.\]
\end{quote}

\begin{quote}
\begin{quote}
The key thing to notice in (6) is that the ratio of residual
variances--from the original model predictions,
\(var_{res_{baseline}}\), and the Shapley-modified predictions,
\(var_{res_{shap}}^{(f)}\)--control the size of \(R_{shap}^{2(f)}\).
This ratio ranges from 1 to approaching 0. When the ratio is 1, removing
the feature does not increase the model's prediction error--equivalent
to a Shapley value of 0 across all instances--and \(R_{shap}^{2(f)}\) =
0. In edge cases where removing a feature counterintuitively leads to
better predictions and reduces the residual variance (e.g., when using
stochastic Shapley value approximations), this ratio should be set to 1
to avoid negative \(R^{2}\) values. The denominator of (6) normalizes
the \(R^{2}\) values to produce a simplex similar to that produced by
the softmax function. Finally, multiplying this vector of feature-level
variance explained attributions with \(R^{2}_{baseline}\) ensures that
\(R^{2}_{baseline} := \sum_{f=1}^{F} R_{shap}^{2(f)}\).
\end{quote}
\end{quote}

One positive byproduct of our approach is that the use of Shapley values
in (5) helps us avoid refitting computationally demanding machine
learning models to capture the unique contribution of each feature. A
model need only be fit once. For our metric, the computational burden is
shifted from model {[}re{]}training to a pre-computed Shapley value
calculation. And while model-agnostic stochastic Shapley value
calculations can be time-intensive themselves {[}5{]}, recent
advancements in interpretable machine learning with gradient boosted
decision trees {[}17{]} and neural networks {[}3{]} support Shapley
value calculations with marginal overhead in medium-sized datasets.

\hypertarget{accounting-for-correlation}{%
\subsection{Accounting for
Correlation}\label{accounting-for-correlation}}

In the case of models with uncorrelated features, each feature in the
proposed \(R^{2}\) decomposition explains a unique proportion of the
variance in the modeled outcome. This is the ideal result and one that
is common across alternative \(R^{2}\) decompositions {[}8{]}. In
practice, modeled features are rarely, if ever, uncorrelated in an
applied problem where machine learning algorithms are considered. The
effect of these non-zero feature correlations on variance-based measures
of feature importance is that, as correlations between features
increase, (a) the unique variance in an outcome explained by each
feature decreases and (b) the remaining model-explained variance is
attributed to the collection of features as a whole. Any model-explained
variance that is aggregated and shared across features is undesirable
when making claims about feature importance--the threat to validity
being that, if this shared explained variance \emph{could} be assigned
to the appropriate features, our feature importance rankings may
qualitatively change.

To be clear, our metric makes claims of global feature importance by
focusing on the residual variance that \emph{can} be uniquely ascribed
to each feature. We can measure this variance, \(\sigma_{unique}\), by
iterating through our features, removing the marginal effect of each
feature by calculating the Shapley-modified prediction shown in (5),
measuring the increase in residual, \(y - \hat{y}^{(f)}_{shap}\),
variance of the predictions resulting from the feature's removal,
summing this increase in residual variance across features, and
computing the ratio of this variance to the residual variance from the
full model with all features. This series of steps can be summarized as

\[\sigma_{unique} = \dfrac{\sum_{f=1}^{F} var(y - \hat{y}^{(f)}_{shap})}
{var(y - \bar{y}) - var(y - \hat{y})} \tag{7}.\]

With uncorrelated features, \(\sigma_{unique}\) equals 1 and the
model-explained variance can be completely and uniquely decomposed into
feature-level shares. As correlations between features increase, this
ratio approaches 0 and, while the feature rankings from the proposed
\(R^{2}\) metric \emph{may} not change, the metric is based on smaller
proportions of uniquely explained variance and the results may become
unstable and highly sample-dependent. We recommend reporting this ratio
alongside the proposed \(R^{2}\) metric to communicate the robustness of
the feature rankings.

To illustrate the extent to which feature correlations affect our
metric, we simulated data from a linear model with varying levels of
correlation between features and calculated \(\sigma_{unique}\). The
data were simulated from 3-feature multivariate standard normal
distributions with uniform correlations, and (7) was calculated using
the stochastic Shapley value approximations described in {[}3{]}. The
results in Figure 1 suggest that even moderate levels of between-feature
correlation affect the validity of variance-based measures of feature
importance as approximately 50\% of the model-explained variance cannot
be assigned to specific features in a straightforward manner.

\begin{figure}[H]
  \centering
  \includegraphics[width=16cm, height=12cm]{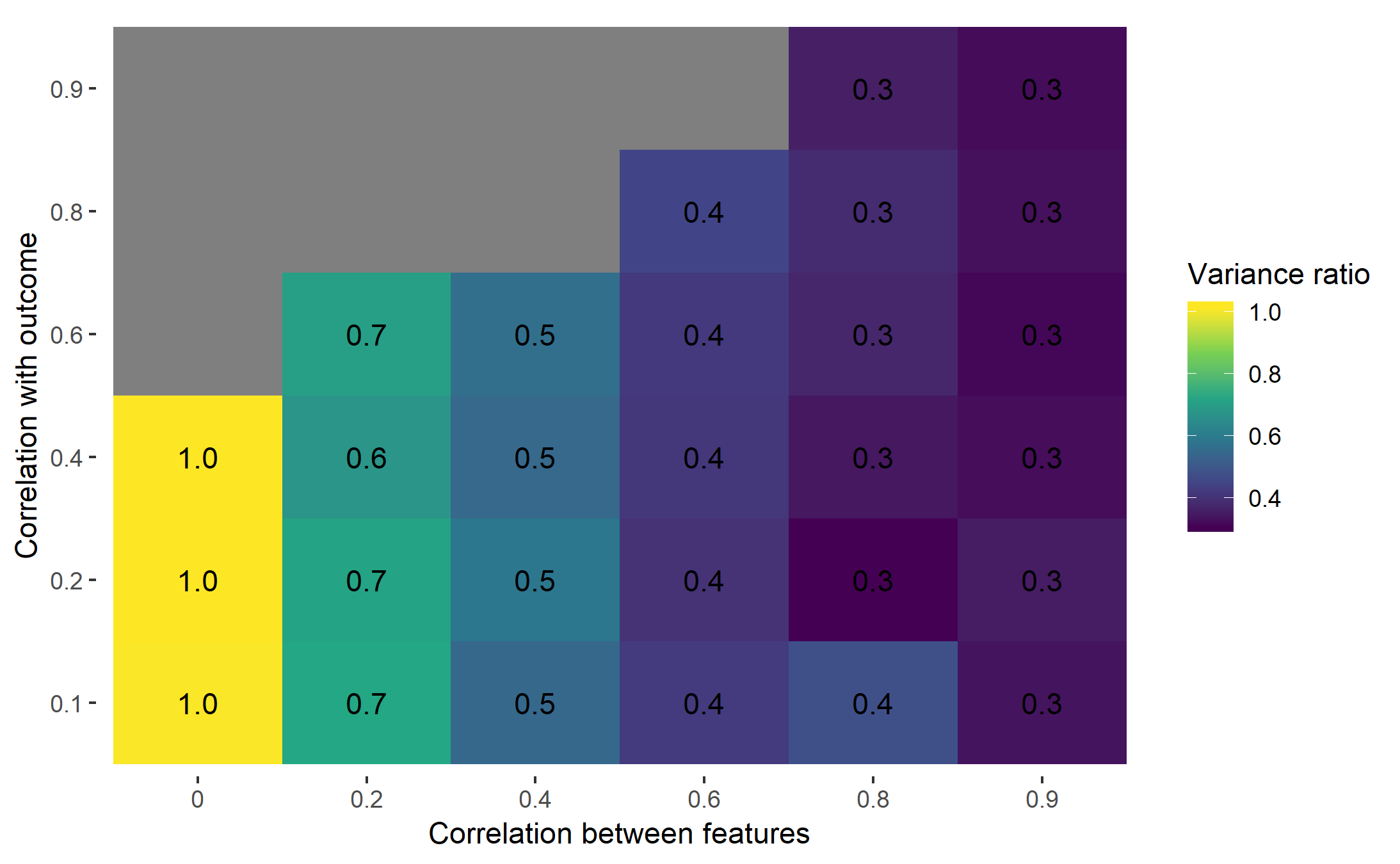}
\caption{
This figure is the result of a simulation that depicts the proportion of model explained variance 
that can be decomposed and uniquely assigned to model 
features, $\sigma_{unique}$, as a function of between-feature correlations and feature relationship with the model outcome. 
The gray boxes indicate simulations that were not run due to non-positive definite correlation matrices.
}
\end{figure}

Recent research from {[}18{]} into kernel-based solutions to the feature
correlation problem in Shapley value analysis holds promise. We do not,
however, cover their approach here because it does not alter the
proposed \(R^{2}\) decomposition; rather, the correlation-adjusted
Shapley values that they propose can be thought of as a data
preprocessing step that strengthens the validity of our metric.

\hypertarget{applied-analysis}{%
\subsection{Applied Analysis}\label{applied-analysis}}

To get a sense of the stability of the proposed \(R^2\) feature
importance metric in an applied setting, we fit a series of gradient
boosted decision tree models to predict (a) wine quality in the white
wine dataset {[}19{]} and (b) the area burned in a forest fire in the
forest fires dataset {[}20{]}, both of which are available on the UCI
Machine Learning Repository {[}21{]}. The models were fit with the
\(\texttt{R}\) implementation of the \(\texttt{catboost}\) package
{[}22{]} version 0.14.2. \(\texttt{catboost}\) was chosen in part
because of its built-in support for the exact Shapley value
calculations, titled Tree SHAP, described in {[}17{]} which are
computationally inexpensive. To assess the metric's stability, a series
of five models were fit to achieve \(R^2\) values ranging from .05 to
.50 using (4). The models were fit using the default settings and
increasing global \(R^2\) values were obtained by increasing the number
of iterations or trees in the ensemble from (a) \textasciitilde{}30 to
\textasciitilde{}3,000 in the white wine dataset and (b)
\textasciitilde{}300 to \textasciitilde{}9,000 in the forest fires
dataset. The feature-level \(R^2\) values were calculated for the
training data--all available data--using our implementation of (6)
available in the \(\texttt{r2()}\) function in the open-source
\(\texttt{R}\) package \(\texttt{shapFlex}\)
(\url{https://github.com/nredell/shapFlex}).

While far short of a thorough simulation, the results in Figure 2
suggest that the variance explained by each feature under the proposed
metric is potentially robust in applied analyses and likely changes
predictably across the spectrum of underfitting to overfitting.
Interestingly, the \(\sigma_{unique}\) proportion of model-explained
variance that could be uniquely assigned to each feature was similar
across datasets: In the wine quality dataset, \(\sigma_{unique}\) ranged
from .88 in the \(R^2\) = .05 model to .68 in the \(R^2\) = .50 model,
and in the forest fires dataset, \(\sigma_{unique}\) ranged from .86 in
the \(R^2\) = .05 model to .67 in the \(R^2\) = .50 model. The
similarity of the\\
size of the \(\sigma_{unique}\) values is likely due to the
near-identical average absolute correlation between numeric features in
each dataset--.17. The similarity in the direction of the
\(\sigma_{unique}\) values--decreasing as the boosted tree models were
increasingly overfit--is more difficult to explain. One plausible
explanation is that overfitting a non-linear model to the data
increasingly captures linear and non-linear relationships in the data
that\\
increasingly violate the feature independence assumption in most Shapley
value and Shapley value approximation algorithms. To what extent this is
data dependent, model dependent, or Shapley value calculation dependent
remains, to our knowledge, an open research question.

\begin{figure}[H]
  \centering
  \includegraphics[width=16cm, height=12cm]{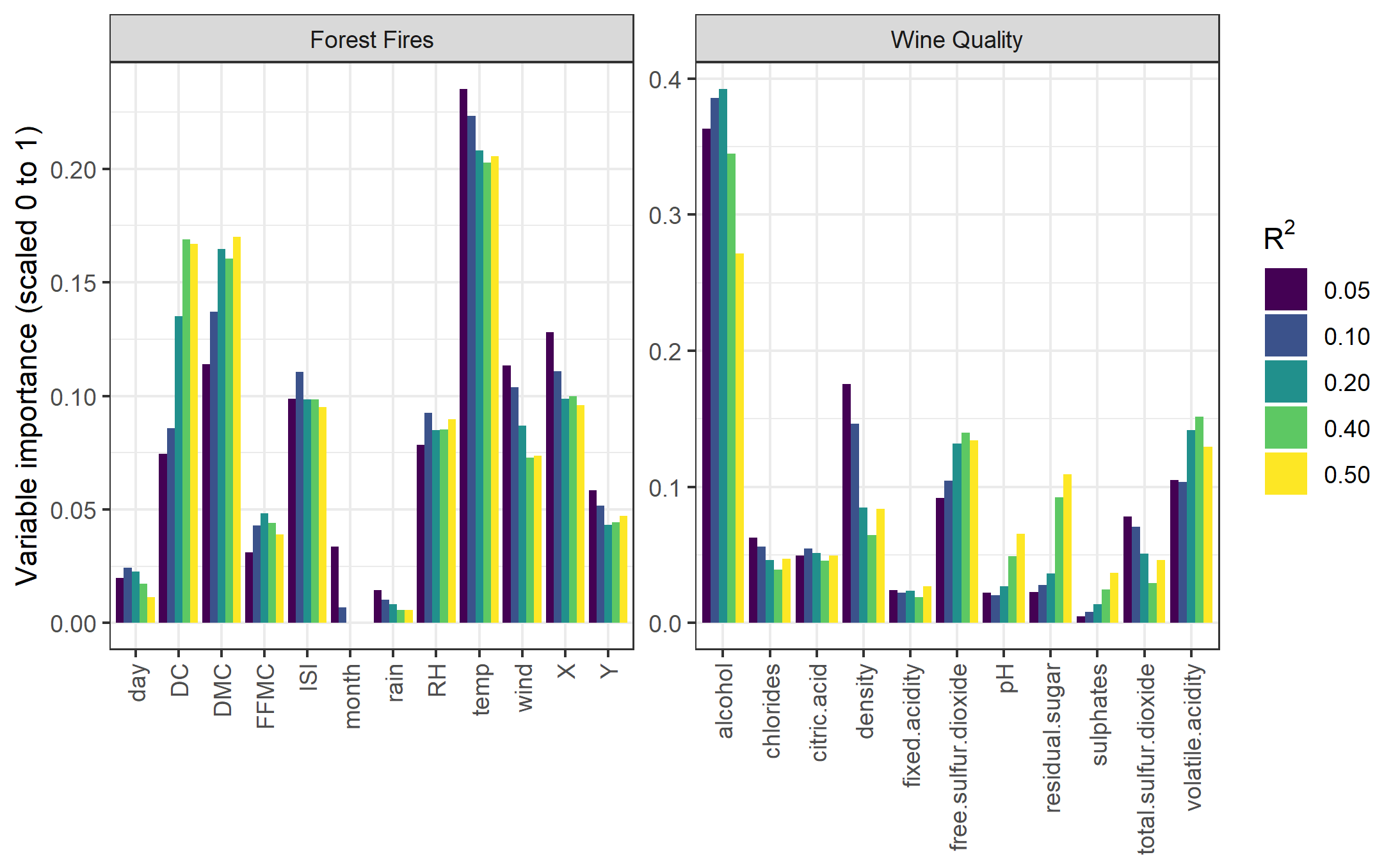}
\caption{This figure depicts the feature-level attribution of the total variance explained in (a) wine quality ratings and (b) the area 
burned in forest fires for each of five models that increasingly explain larger proportions of the observed variance in the outcomes. 
The results have been scaled from 0 to 1, as opposed to each model's $R^2$, to facilitate comparisions across models.}
\end{figure}

\hypertarget{discussion}{%
\subsection{Discussion}\label{discussion}}

In this paper we proposed a modified version of \(R^2\) suitable for
summarizing feature importance in linear and non-linear machine learning
models. This metric makes use of the additive property of Shapley values
to fairly distribute the share of a model's explained variance to each
feature without the computational burden of refitting a machine learning
model to various subsets of features. However, while the proposed metric
has nice properties--0 to 1 scaling and a feature-level variance
decomposition summing to the overall model \(R^2\)--, the effect of
overfitting and hyperparameter tuning on the stability of the
decomposition needs further investigation. Additional open questions
include whether this metric is best applied to model training or model
testing data as well as how this metric could be extended to
hierarchical or nested data structures where explaining the variability
between groups or through time is of interest. Nonetheless, the proposed
\(R^2\) metric could see use by practitioners looking to better
understand the models that they have built by bridging the gap between
classical statistics and modern machine learning.

\hypertarget{references}{%
\subsection{References}\label{references}}

\setlength{\parindent}{-0.5in}
\setlength{\leftskip}{0.5in}
\setlength{\parskip}{8pt}

\noindent

{[}1{]} Murdoch, W. J., Singh, C., Kumbier, K., Abbasi-Asl, R., \& Yu,
B. (2019). Interpretable machine learning: definitions, methods, and
applications. arXiv preprint arXiv:1901.04592.

{[}2{]} Doshi-Velez, F., \& Kim, B. (2017). Towards a rigorous science
of interpretable machine learning. arXiv preprint arXiv:1702.08608.

{[}3{]} Lundberg, S. M., \& Lee, S. I. (2017). A unified approach to
interpreting model predictions. In Advances in Neural Information
Processing Systems (pp.~4765-4774).

{[}4{]} Biecek, P. (2018). DALEX: explainers for complex predictive
models in R. The Journal of Machine Learning Research, 19(1), 3245-3249.

{[}5{]} Molnar, C., Casalicchio, G., \& Bischl, B. (2018). Iml: An R
package for interpretable machine learning. The Journal of Open Source
Software, 3(786), 10-21105.

{[}6{]} Molnar, Christoph. (2019) ``Interpretable machine learning. A
Guide for Making Black Box Models Explainable''.
\url{https://christophm.github.io/interpretable-ml-book/}.

{[}7{]} Britton, M. (2019). VINE: Visualizing Statistical Interactions
in Black Box Models. arXiv preprint arXiv:1904.00561.

{[}8{]} Grömping, U. (2015). Variable importance in regression models.
Wiley Interdisciplinary Reviews: Computational Statistics, 7(2),
137-152.

{[}9{]} Wachter, S., Mittelstadt, B., \& Russell, C. (2017).
Counterfactual explanations without opening the black box: Automated
decisions and the GDPR. Harvard Journal of Law \& Technology, 31(2),
2018.

{[}10{]} Wright, S. (1921). Correlation and causation. Journal of
agricultural research, 20(7), 557-585.

{[}11{]} Kvålseth, T. O. (1985). Cautionary note about R 2. The American
Statistician, 39(4), 279-285.

{[}12{]} Fisher, A., Rudin, C., \& Dominici, F. (2018). All Models are
Wrong but many are Useful: Variable Importance for Black-Box,
Proprietary, or Misspecified Prediction Models, using Model Class
Reliance. arXiv preprint arXiv:1801.01489.

{[}13{]} Gelman, A., Goodrich, B., Gabry, J., \& Vehtari, A. (2018).
R-squared for Bayesian regression models. The American Statistician,
(just-accepted), 1-6.

{[}14{]} Lideman, R., Merenda, P., \& Gold, R. (1980). Introduction to
bivariate and multivariate analysis scott. Scott Foresman: Glenview, IL,
USA.

{[}15{]} Coleman, C. D., (2017) Decomposing the R-squared of a
Regression Using the Shapley Value in SAS ®. US Census Bureau.

{[}16{]} Shapley, L. S. (1953). A value for n-person games. In Kuhn, H.
W. and Tucker, A. W., editors, Contribution to the Theory of Games II
(Annals of Mathematics Studies 28), pages 307--317. Princeton University
Press, Princeton, NJ.

{[}17{]} Lundberg, S. M., Erion, G. G., \& Lee, S. I. (2018). Consistent
individualized feature attribution for tree ensembles. arXiv preprint
arXiv:1802.03888.

{[}18{]} Kjersti, Aas, Jullum, Martin, \& Løland, Anders (2019).
Explaining individual predictions when features are dependent: More
accurate approximations to Shapley values. arXiv preprint
arXiv:1903.10464v2.

{[}19{]} P. Cortez, A. Cerdeira, F. Almeida, T. Matos and J. Reis.
(2009). Modeling wine preferences by data mining from physicochemical
properties. In Decision Support Systems, Elsevier, 47(4):547-553.

{[}20{]} P. Cortez \& A. Morais. (2007). A Data Mining Approach to
Predict Forest Fires using Meteorological Data. In J. Neves, M. F.
Santos and J. Machado Eds., New Trends in Artificial Intelligence,
Proceedings of the 13th EPIA 2007 - Portuguese Conference on Artificial
Intelligence, December, Guimarães, Portugal, pp.~512-523, 2007. APPIA,
ISBN-13 978-989-95618-0-9.

{[}21{]} Dua, D. \& Graff, C. (2019). UCI Machine Learning Repository
{[}\url{http://archive.ics.uci.edu/ml}{]}. Irvine, CA: University of
California, School of Information and Computer Science.

{[}22{]} Prokhorenkova, L., Gusev, G., Vorobev, A., Dorogush, A. V., \&
Gulin, A. (2018). CatBoost: unbiased boosting with categorical features.
In Advances in Neural Information Processing Systems (pp.~6638-6648).

\end{document}